\documentclass[a4paper,10pt,twoside]{cpc-hepnp}
\usepackage{CJK,upgreek,fancyhdr}
\usepackage{multicol}
\usepackage{booktabs}
\usepackage{amssymb,bm,mathrsfs,bbm,amscd}
\usepackage[tbtags]{amsmath}
\usepackage{lastpage}
\usepackage[T1]{fontenc}
\usepackage{float}
\usepackage{amsbsy}
\usepackage{amsmath}
\usepackage{amstext}
\usepackage{hyperref}
\usepackage{graphicx}
\usepackage{esint}
\usepackage{subfigure}
\usepackage{upgreek}
\usepackage{wrapfig}
\usepackage{amscd}
\usepackage{xcolor}
\usepackage{hyperref}
\usepackage{overpic}
\usepackage{multirow}
\usepackage{color}
\usepackage{graphicx}

\usepackage[latin9]{inputenc}
\usepackage{array}

\graphicspath{{pic/}}
\begin{document}
\begin{CJK*}{GBK}{song}

\fancyhead[c]{\small Submitted to Chinese Physics C}
\fancyfoot[C]{\small 010201-\thepage}


\title{Evaluating topological charge density with symmetric multi-probing
method\thanks{Supported in part by the National Natural Science Foundation of China (NSFC) under the project No.11335001, No.11275169.} }

\author{%
      Guang-Yi Xiong $^{\dagger;1}$\email{xionggy@zju.edu.cn}%
\quad Jian-Bo Zhang $^{\dagger;2}$\email{jbzhang08@zju.edu.cn}%
\quad You-Hao Zou $^{\dagger;3}$\email{11006067@zju.edu.cn}%
\\
}
\maketitle

\address{%
$^\dagger$ Department of Physics, Zhejiang University, Zhejiang 310027,
P.R. China\\
}

\begin{abstract}
We evaluated the topological charge density of SU(3) gauge fields on lattice
by calculating the trace of overlap Dirac matrix employing symmetric multi-probing(SMP)
method with 3 modes. {\color{black}Since} the topological charge $Q$ for a given lattice configuration
must be an integer number, {\color{black}it's}
easy to estimate the systematic error (the deviation of $Q$ to nearest
integer). The results {\color{black}showed} high efficiency and accuracy in calculating
the trace of the inverse of a large sparse matrix with locality by
using SMP sources, compared with that using point sources. We also {\color{black}showed}
the correlation between the errors and probing scheme parameter
$r_{\mathrm{min}}$ as well as lattice volume $N_{L}$ and lattice spacing
$a$. It {\color{black}was} found that the computing time of calculating
the trace by employing SMP sources is less dependent on $N_{L}$
than that by using point sources. Therefore the SMP method is very suitable
for calculations on large lattices.
\end{abstract}



\footnotetext[0]{\hspace*{-3mm}\raisebox{0.3ex}{$\scriptstyle\copyright$}2018
Chinese Physical Society and the Institute of High Energy Physics
of the Chinese Academy of Sciences and the Institute
of Modern Physics of the Chinese Academy of Sciences and IOP Publishing Ltd}%

\section{introduction}
Topological charge $Q$ and density $q\left(x\right)$ are important
quantities to the investigation of the structure of quantum chromodynamics(QCD) vacuum, and the
susceptibility of $Q$ is related to $\eta^{\prime}$
mass by Witten-Veneziano relation~\cite{witten1979current,veneziano1979u}. Lattice QCD is one of
the best non-perturbative framework to study QCD properties, in which space-time is discretized on lattice.
$Q$ on lattice can be calculated by gauge field tensor $F_{\mu\nu}$~\cite{luscher1982topology}:
\begin{equation}
Q=\int\mathrm{d}^{4}xq\left(x\right)=\frac{1}{32\pi^{2}}\int\mathrm{d}^{4}x\mathrm{tr}\left(\epsilon_{\mu\nu\rho\sigma}F_{\mu\nu}F_{\rho\sigma}\right),
\end{equation}
where the trace is over color indexes, and $\epsilon_{\mu\nu\rho\sigma}$
is a totally antisymmetric tensor. However, the topological
charge $Q$ calculated in this way is always not an integer unlike that
in the continuum, unless multiplied by a renormalization factor
or calculated on smoothed configurations whose renormalization
factor tends to 1. Smoothing methods {\color{black}or other cooling-like methods}
will change the topological density,
and also introduce ambiguous smoothing parameter like smoothing time
$n_{\mathrm{sm}}$ which is always set empirically. Another way to
get $q\left(x\right)$ is evaluating the trace of chiral
Dirac fermion operator in Dirac and color space~\cite{Niedermayer:1998bi}:
\begin{equation}
q\left(x\right)=\frac{1}{2}\mathrm{tr}\left(\gamma_{5}D_{\mathrm{chiral}}\right),
\end{equation}
where the trace is over color and Dirac indexes, $\gamma_{5}$ is
Dirac gamma matrix, and $D_{\mathrm{chiral}}$ is a chiral Dirac fermion
operator on lattice. In this definition, $Q$ will be always an integer due to
Atiya-Singer index theorem~\cite{atiyah1971index},
but hard to be exactly calculated as evaluating the trace of $D_{\mathrm{chiral}}$
is a very expensive task. As one of the chiral Dirac fermions, overlap fermion operator~\cite{Neuberger:1997fp,Neuberger:1998wv}
is widely used in the studies of chiral properties of lattice QCD. Massless
overlap Dirac operator $D_{\mathrm{ov}}$ is constructed by a kernal
operator, usually using Wilson Dirac operator $D_{\mathrm{W}}$ as
the kernal:
\begin{equation}
D_{\mathrm{ov}}=1+\gamma_{5}\epsilon\left(\gamma_{5}D_{\mathrm{W}}\right),\label{eq:Dov}
\end{equation}
where the hopping parameter $\kappa$ in the kernal $D_{\mathrm{W}}$
is only a free parameter not related to bare quark mass, and $\kappa\in\left(\kappa_{c},0.25\right)$
with $\kappa_{c}$ refer to critical hopping parameter in Wilson fermion. We {\color{black}took} $\kappa=0.21$
in this work. $\epsilon\left(\gamma_{5}D_{W}\right)$ is the sign function here:
\begin{equation}
\epsilon(\gamma_{5}D_{\mathrm{W}})=\mathrm{sign}\left(\gamma_{5}D_{\mathrm{W}}\right)=\frac{\gamma_{5}D_{\mathrm{W}}}{\sqrt{D_{\mathrm{W}}^{\dagger}D_{\mathrm{W}}}}=\gamma_{5}D_{W}\left(c_{0}+\sum_{i=1}^{\infty}\frac{c_{i}}{D_{\mathrm{W}}^{\dagger}D_{\mathrm{W}}+q_{i}}\right),\label{eq:Zolo}
\end{equation}
in which Zolotarev series expansion~\cite{zolotarev1877application}
is introduced. In our work 14th-order Zolotarev expansion was applied
~\cite{chen2004chiral}. {\color{black}To extract} the trace of $D_{\mathrm{ov}}${\color{black}, we
need} to calculate the trace of $D_{\mathrm{W}}^{-2}$, while $D_{\mathrm{W}}$
is a large sparse matrix. Its rank is $12N_{L}$,
where $N_{L}$ is the number of lattice sites, empirically in a range of
$10^{5}\sim10^{8}$ depending on the lattice size. So the trace of $D_{\mathrm{W}}^{-2}$
is too expensive to calculate exactly.

Fortunately $D_{\mathrm{ov}}$ is a local operator~\cite{Hernandez:1998et},
and there are methods to approximately calculate the trace of it by making
use of this property. Multi-probing method is one of these methods
~\cite{Bekas2007An,Tang2012} that can be used to estimating the trace
of the inverse of a large sparse matrix when the inverse exhibits
locality. A. Stathopoulos et al. developed the hierarchical multi-probing
method and tested it in lattice QCD Monte Carlo calculation~\cite{stathopoulos2013hierarchical}.
E. Endress et al. used multi-probing method to tackle the noise-to-signal
problems when calculating all-to-all propagator in lattice QCD~\cite{Endress:2014qpa,Endress:2014ppa}
by applying the Greedy Multicoloring Algorithm to construct the multi-probing
source vectors. Their results have showed the validity and efficiency
of probing method.

In this work we {\color{black}employed} symmetric multi-probing(SMP) method to
evaluating the topological charge density by calculating $\mathrm{tr}\left(\gamma_{5}D_{\mathrm{ov}}\right)$,
and {\color{black}showed} its efficiency comparing to point source method.
We will introduce the topological charge density as well as the trace
calculation with point sources and SMP sources in
next section. After that, calculation results will be presented and
discussed. We will summarize our work in the last section.

\section{methods}

\subsection{Topological charge density and trace calculation with point source }

Overlap Dirac operator $D_{\mathrm{ov}}\left(x,y\right)_{\alpha\beta,ab}$
on a lattice $L(n_{x},n_{y},n_{t},n_{z})$ is a large sparse matrix,
where $n_{\mu}$ denotes the number of sites in $\mu$ direction and
$\mu=x,y,z,t$, $\alpha$ $\beta$ are Dirac indexes and $a$ $b$ are
color indexes. The topological charge density $q\left(x\right)$
can be obtained by calculating the trace of massless overlap Dirac operator:
\begin{equation}
q(x)=\frac{1}{2}\mathrm{tr}_{\mathrm{d,c}}\left(\gamma_{5}D_{\mathrm{ov}}\left(x,x\right)\right)=\frac{1}{2}\sum_{\alpha,a}\left\langle \psi\left(x,\alpha,a\right)\right|\gamma_{5}D_{\mathrm{ov}}\left|\psi\left(x,\alpha,a\right)\right\rangle ,\label{eq:qx}
\end{equation}
where the trace is over Dirac and color indexes, $|\psi(x,\alpha,a)\rangle$
is a normalized point source vector which has $N_{L}\times N_{d}\times N_{c}$
components where the lattice volume $N_{L}=n_{x}n_{y}n_{z}n_{t}$,
$N_{d}=4$, $N_{c}=3$. Only one specific
$\left(x,\alpha,a\right)$ component takes nonzero value in the point
source. Substituting Eq. (\ref{eq:Dov}) and Eq. (\ref{eq:Zolo})
into Eq. (\ref{eq:qx}), the most expensive step to calculate the
trace of $D_{\mathrm{ov}}$ is computing the multiplication:
\begin{equation}
v_{i}=\frac{c_{i}}{D_{\mathrm{W}}^{\dagger}D_{\mathrm{W}}+q_{i}}\left|\psi\left(x,\alpha,a\right)\right\rangle ,
\end{equation}
which is equivalent to solve the following linear equation:
\begin{equation}
M_{i}v_{i}=b,\label{eq:Mx=00003Db}
\end{equation}
where
\begin{equation}
M_{i}=\frac{1}{c_{i}}\left(D_{\mathrm{W}}^{\dagger}D_{\mathrm{W}}+q_{i}\right),\ b=\left|\psi\left(x,\alpha,a\right)\right\rangle .
\end{equation}
This equation is expensive to solve for a large {\color{black}matrix $M_{i}$}, and the huge number
of point sources corresponding to sites on the whole lattice makes things
even worse. The computing time to solve Eq. (\ref{eq:Mx=00003Db})
with Conjugated Gradient(CG) algorithm for one source is more than
$\mathcal{O}\left(N_{L}\right)$ (in consideration of that $M_{i}$ is a
large sparse matrix, and its conditional number grows as
 $N_{L}$ grows), and there are $12N_{L}$ point sources.
So the cost of evaluating $Q=\sum_{x}q\left(x\right)$ by calculating
the trace of $D_{\mathrm{ov}}$ with point sources will be larger
than $\mathcal{O}\left(N_{L}^{2}\right)$, which is too expensive
especially for large $N_{L}$~\cite{Ilgenfritz:2007xu,Koma:2010vx},
although it's the exact measurement of topological charge density
on lattice without any approximation.

\subsection{Trace calculation with multi-probing source}

\begin{figure}[h t b]
\begin{centering}
\includegraphics[scale=0.4]{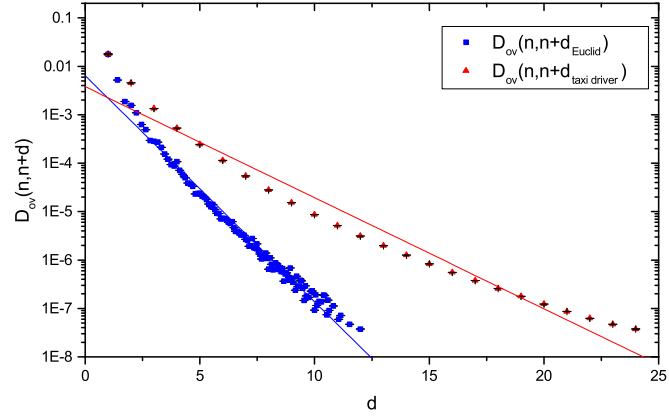}
\par\end{centering}
\caption{The locality of $D_{\mathrm{ov}}$ is shown by its averaged off-diagonal elements as a function of Euclid distance and taxi driver distance with logarithmic coordinates. Red line fits the taxi driver distance function with the slope $-0.528$, while blue line fits the Euclid distance function with the slope $-1.075$ in a lattice configuration.}
\label{Fig:Locality}
\end{figure}

We know from above that calculation of the trace of $D_{\mathrm{ov}}$ with point {\color{black}source}
is a quite expensive task. Introduce a new source vector $\phi$ by adding
two point sources:
\begin{equation}
\phi\left(x,y,\alpha,a\right)=\psi\left(x,\alpha,a\right)+\psi\left(y,\alpha,a\right),\ x\neq y,
\end{equation}
then Eq. (\ref{eq:qx}) can be rewritten as:
\begin{eqnarray}
q(x) & = & \sum_{\alpha,a}\psi\left(x,\alpha,a\right)\tilde{D}_{\mathrm{ov}}\phi\left(x,y,\alpha,a\right)\nonumber \\
 &  & -\sum_{\alpha,a}\psi\left(x,\alpha,a\right)\tilde{D}_{\mathrm{ov}}\psi\left(y,\alpha,a\right),
\end{eqnarray}
where $\tilde{D}_{\mathrm{ov}}=\frac{1}{2}\gamma_{5}D_{\mathrm{ov}}$,
and the second term:
\begin{equation}
\sum_{\alpha,a}\psi\left(x,\alpha,a\right)\tilde{D}_{\mathrm{ov}}\psi\left(y,\alpha,a\right)=\sum_{\alpha,a}\tilde{D}_{\mathrm{ov}}(x,\alpha,a;y,\alpha,a)
\end{equation}
are space-time off-diagonal elements of $D_{\mathrm{ov}}$. Considering
the space-time locality of $D_{\mathrm{ov}}$, it has upper bound~
\cite{Hernandez:1998et,Ilgenfritz:2007xu}:
\begin{equation}
\left|D_{\mathrm{ov}}\left(x,\alpha,a;y,\beta,b\right)\right|\le C\exp\left(-\gamma\left|x-y\right|\right),\label{eq:Dlocal}
\end{equation}
where $C$ and $\gamma$ are constants irrelevant to gauge field configurations, and $\left|x-y\right|$
is defined as the distance between site $x$ and site $y$.
Note that $\gamma_{5}$ only {\color{black}exchanges} the elements of $D_{\mathrm{ov}}$
in Dirac space, so $\tilde{D}_{\mathrm{ov}}$ also satisfies this
inequation. 'off-diagonal' will refer to 'space-time off-diagonal' in the following
unless otherwise stated. We checked this relation on several lattice configurations, as
Fig. \ref{Fig:Locality} shown. $\gamma$ in the inequation (\ref{eq:Dlocal})
is around $1.076$ by~\cite{Ilgenfritz:2007xu} where $\left|x-y\right|$
is defined as Euclid distance, and around $0.529$ {\color{black}as} taxi driver distance. The result means that the off-diagonal
elements of $D_{\mathrm{ov}}$ are exponentially suppressed as $\left|x-y\right|$
increases, and the inequation (\ref{eq:Dlocal}) holds when the distance
is not too small (larger than 3) for two definitions of distance. We can also see that $\gamma$ for Euclid
distance is larger which means stronger locality, so we adopted this definition in our work.

So the approximate topological charge density {\color{black}$q\left(x\right)$} can
be written as:
\begin{eqnarray}
q^{\mathrm{approx}}(x) & = & \sum_{\alpha,a}\psi(x,\alpha,a)\tilde{D}_{\mathrm{ov}}\phi(x,y,\alpha,a),\\
q^{\mathrm{approx}}(y) & = & \sum_{\alpha,a}\psi(y,\alpha,a)\tilde{D}_{\mathrm{ov}}\phi(x,y,\alpha,a).
\end{eqnarray}
We can see that the approximate density of both sites can be easily calculated
from $\tilde{D}_{\mathrm{ov}}\phi(x,y,\alpha,a)$, so only one
linear equation like Eq. (\ref{eq:Mx=00003Db}) need to be solved
to obtain $q\left(x\right)$ and $q\left(y\right)$ as long as $\left|x-y\right|$
is sufficiently large. This method is called as multi-probing method
or probing method.

\begin{figure}[h t b]
\begin{centering}
\includegraphics[scale=0.2]{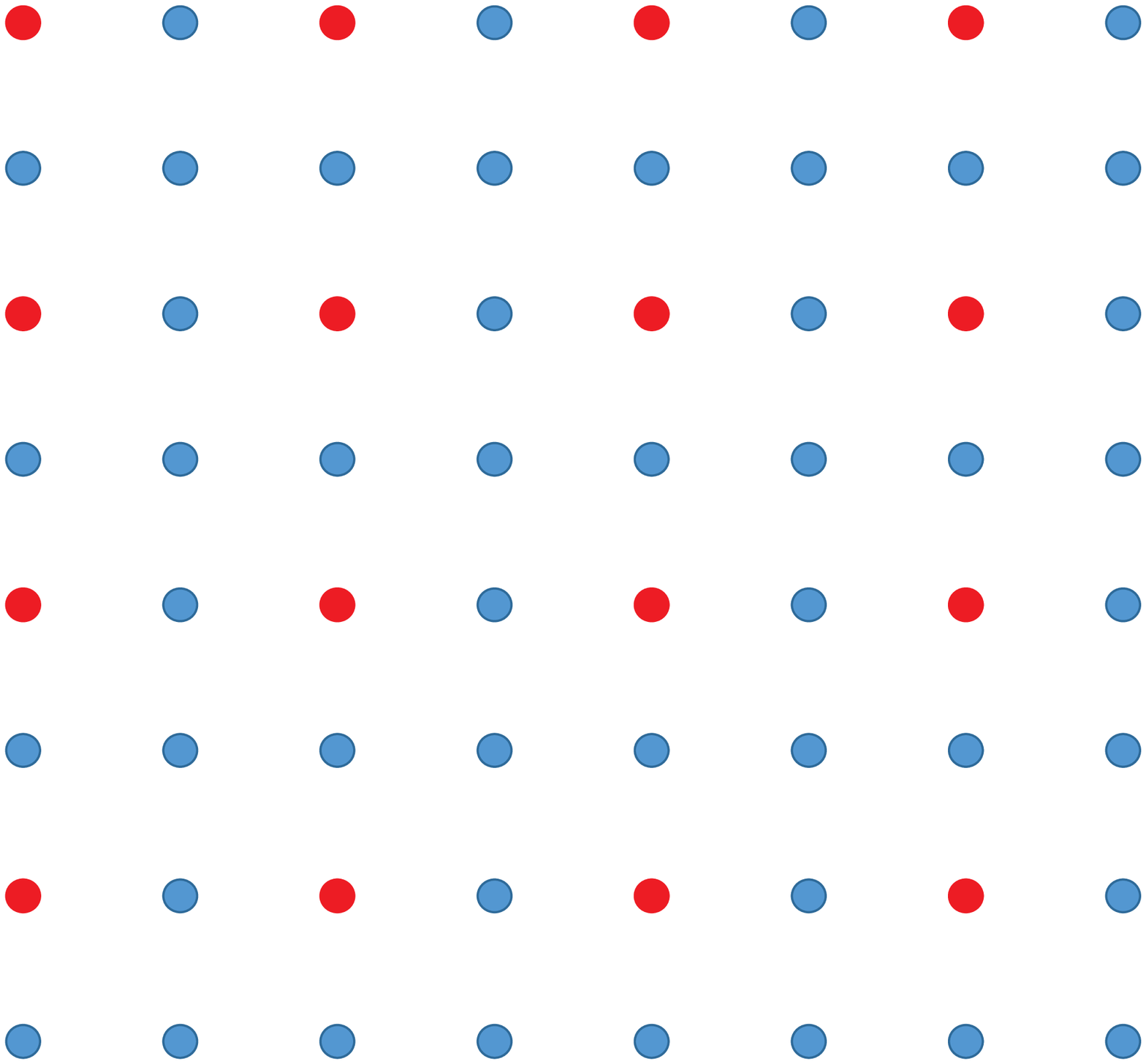} ~~~~~~~~~~~~\includegraphics[scale=0.2]{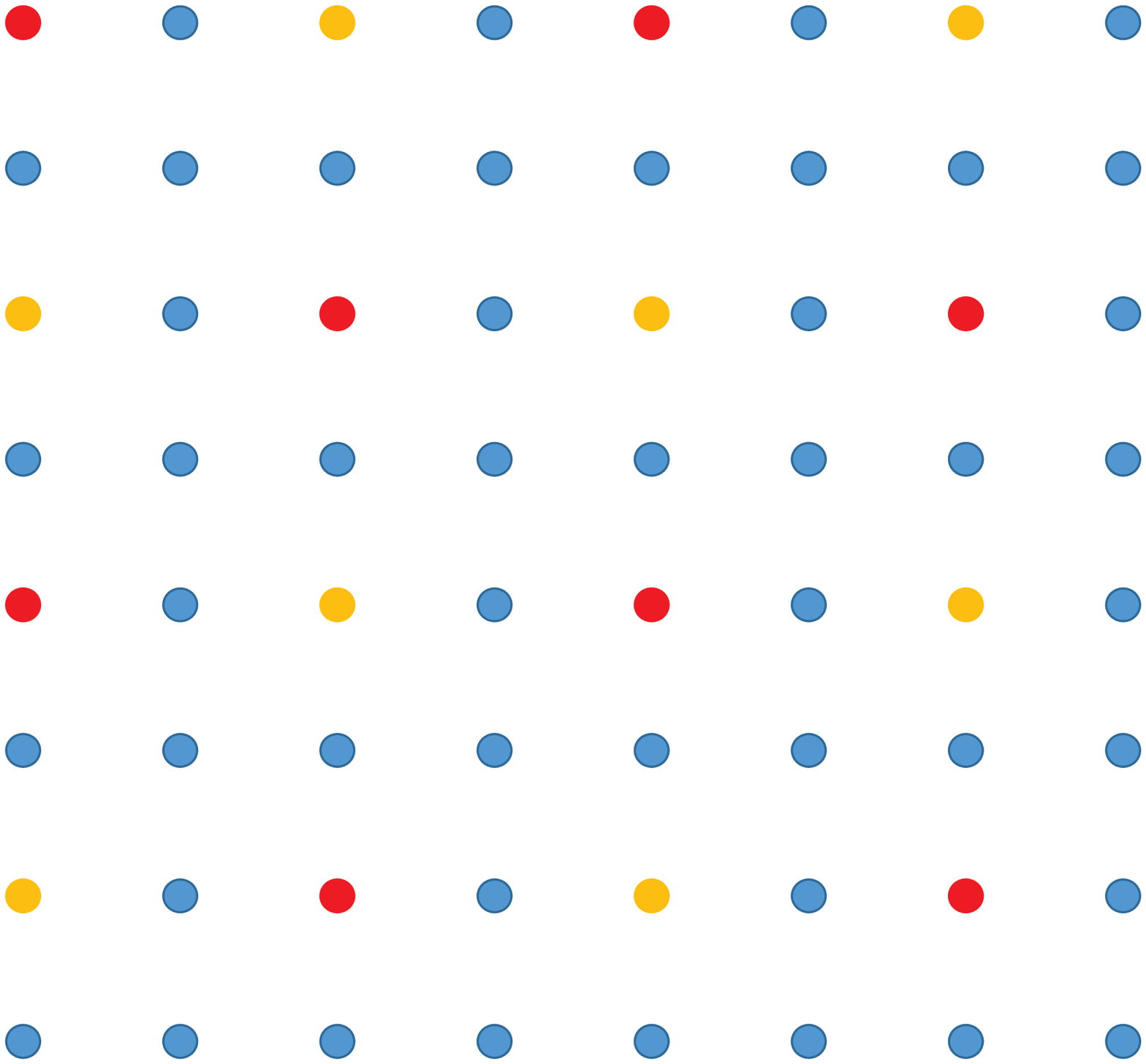}
\par\end{centering}
\caption{Example of Split Mode on $8\times8$ 2-D lattice. Red sites in the left {\color{black}panel} are colored in Normal Mode coloring scheme {\color{black}$P(4,4,0)$}, and in the right {\color{black}panel} they split into two subsets with red and yellow color in Split Mode scheme {\color{black}$P(4,4,1)$}.}
\label{Fig:2D_split}
\end{figure}

\begin{figure}[h t b]
\begin{centering}
\includegraphics[scale=0.2]{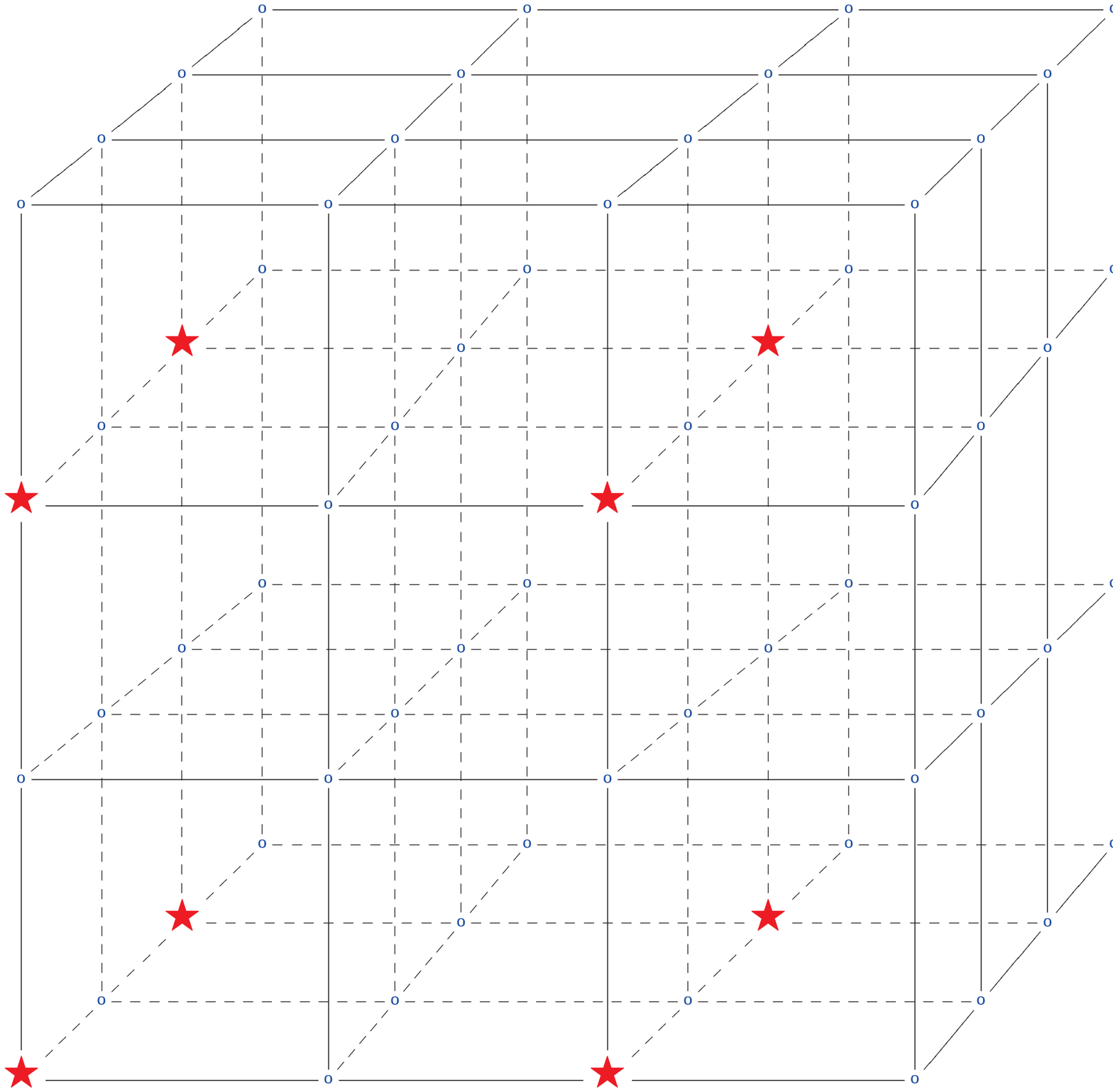} ~~~~~~~~~~~~\includegraphics[scale=0.2]{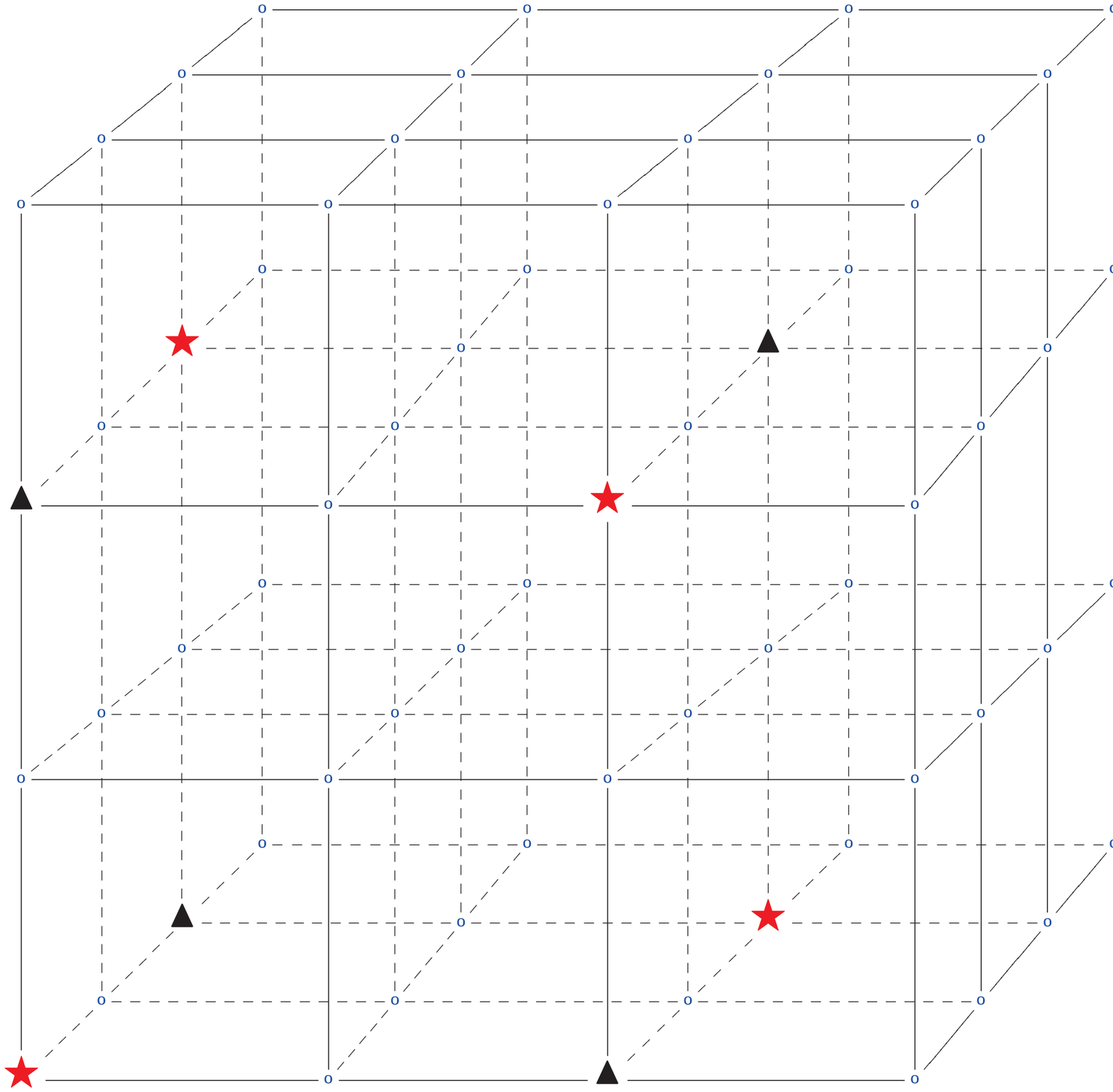}
\par\end{centering}
\caption{Examples of Split Mode on $4^3$ 3-D lattice. sites denoted as red stars in the
left panel are colored in Normal Mode coloring scheme {\color{black}$P(2,2,2,0)$},
while in the right panel {\color{black}they split} into two subsets, red stars
and black triangles in Split Mode {\color{black}scheme $P(2,2,2,1)$}.}
\label{Fig:3D_split}
\end{figure}

{\color{black}We firstly mark} all sites in the lattice with a color label, called coloring algorithm.
The sites with the same color form a subset of the whole lattice sites set. Then one
multi-probing source can be constructed by a subset as follow: choose a color and
Dirac index, construct point sources on sites in the subset and add all these point source vectors.

In order to reduce the calculation cost, we would {\color{black}like} to reduce the number of MP
sources by combining more point sources in one MP source. However, the errors introduced
 by MP source will exponentially depend on the distance between the sites
in the subset, as Eq. (\ref{eq:Dlocal})
shown. We should include the sites that as far {\color{black}away} as possible
from each other in one MP source to control the {\color{black}systematic errors}.
Therefore appropriate coloring
algorithm is important to optimize the balance of the {\color{black}systematic errors} and calculation cost.

We {\color{black}chose} a symmetric coloring algorithm and Euclid distance definition in this work, which
marks the sites with same color in each direction with
fixed distance. There are other algorithms like GCA(Greedy coloring Algorithm) \cite{Tang2012}
who used taxi driver distance definition. {\color{black}It's convenient to} choose the same gap $r$ in each
direction, then the minimal distance of each two sites in the subsets $r_{\mathrm{min}}=r$. We call this type of
MP sources by Normal Mode, as we show some examples in left {\color{black}panel} of Fig. \ref{Fig:2D_split} and
Fig. \ref{Fig:3D_split}.

Furthermore, we can divided a subset into two subsets by marking half of sites with {\color{black}a} new color, while each site
has different color with its neighbors, see examples in Fig. \ref{Fig:2D_split} and Fig. \ref{Fig:3D_split}
on 2-D and 3-D lattices. In this Split Mode, $r_{\mathrm{min}}^{\prime}=\sqrt{2}r$ for the two new {\color{black}subsets}.

\begin{figure}[h t b]
\begin{centering}
\includegraphics[scale=0.2]{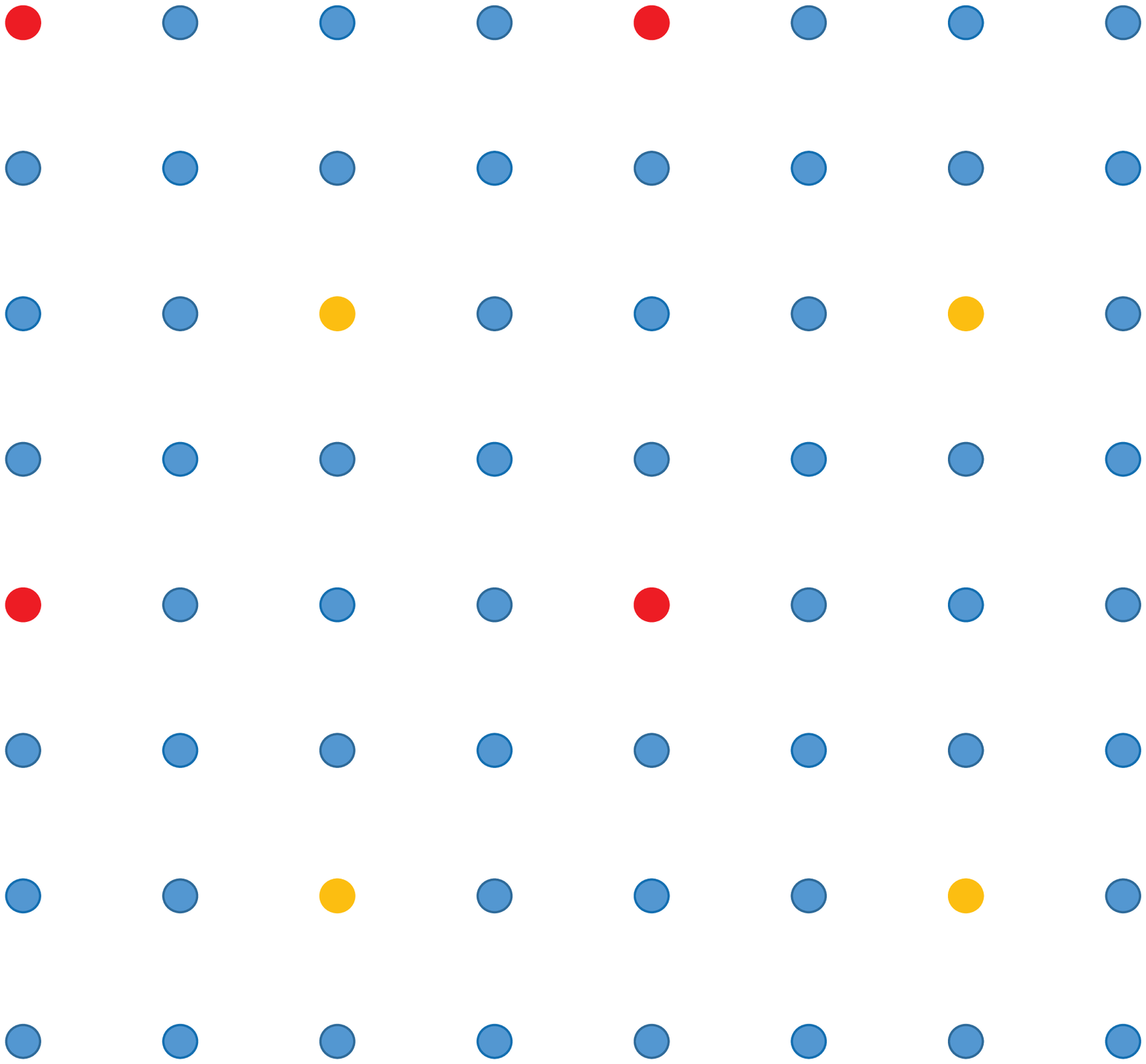} ~~~~~~~~~~~~\includegraphics[scale=0.2]{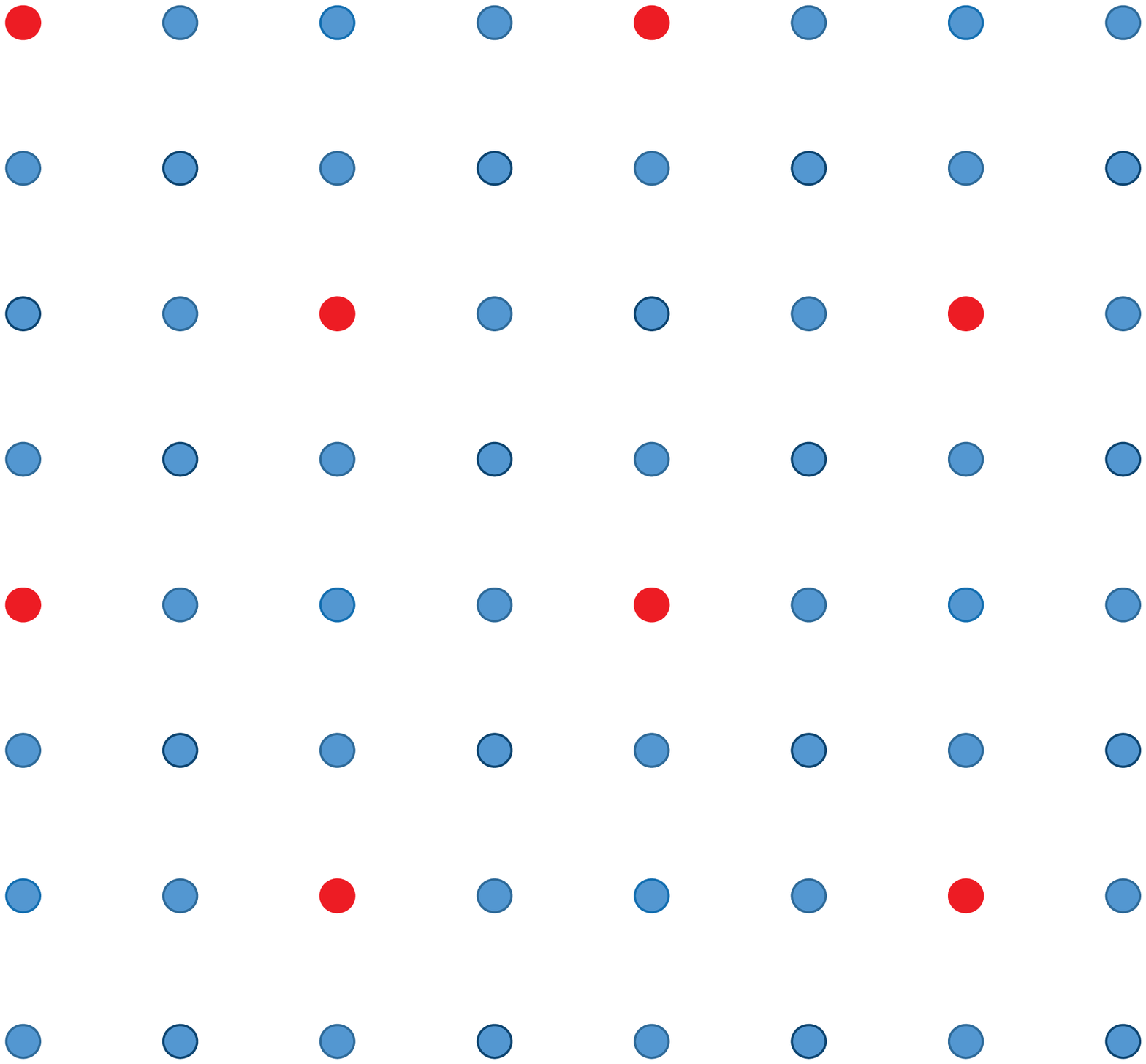}
\par\end{centering}
\caption{Example of combined mode on $8\times8$ 2-D lattice. Red sites and yellow sites in the left {\color{black}panel} are two subsets colored in normal symmetric coloring scheme $P(2,2,0)$, and in the right {\color{black}panel} they are combined into one with red color in combined mode {\color{black}scheme} $P(2,2,2)$. Note that the minimal distance changes {\color{black}from $4$ to $2\sqrt{2}$}.}
\label{Fig:2D_combine}
\end{figure}

\begin{figure}[h t b]
\begin{centering}
\includegraphics[scale=0.2]{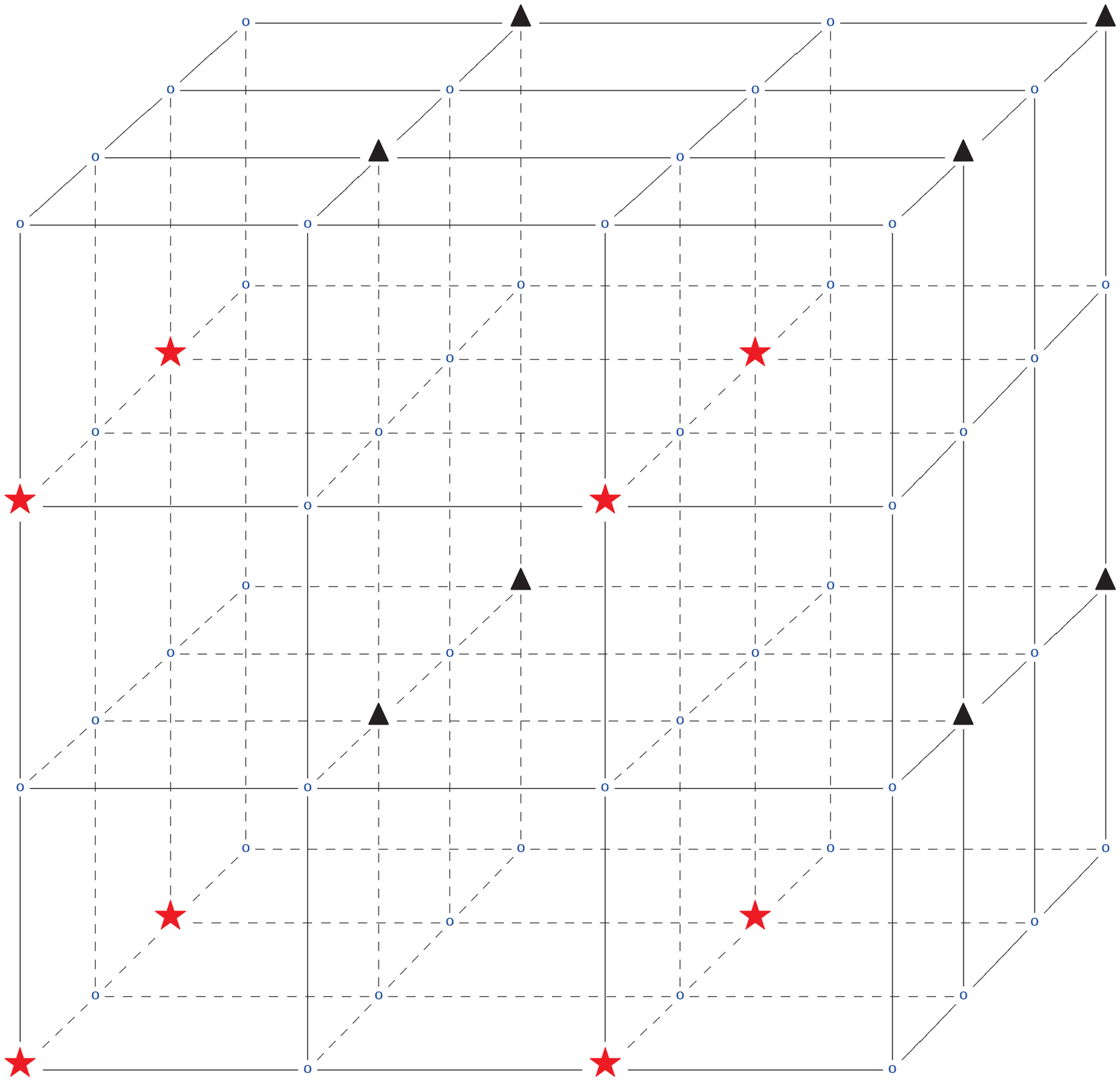} ~~~~~~~~~~~~\includegraphics[scale=0.2]{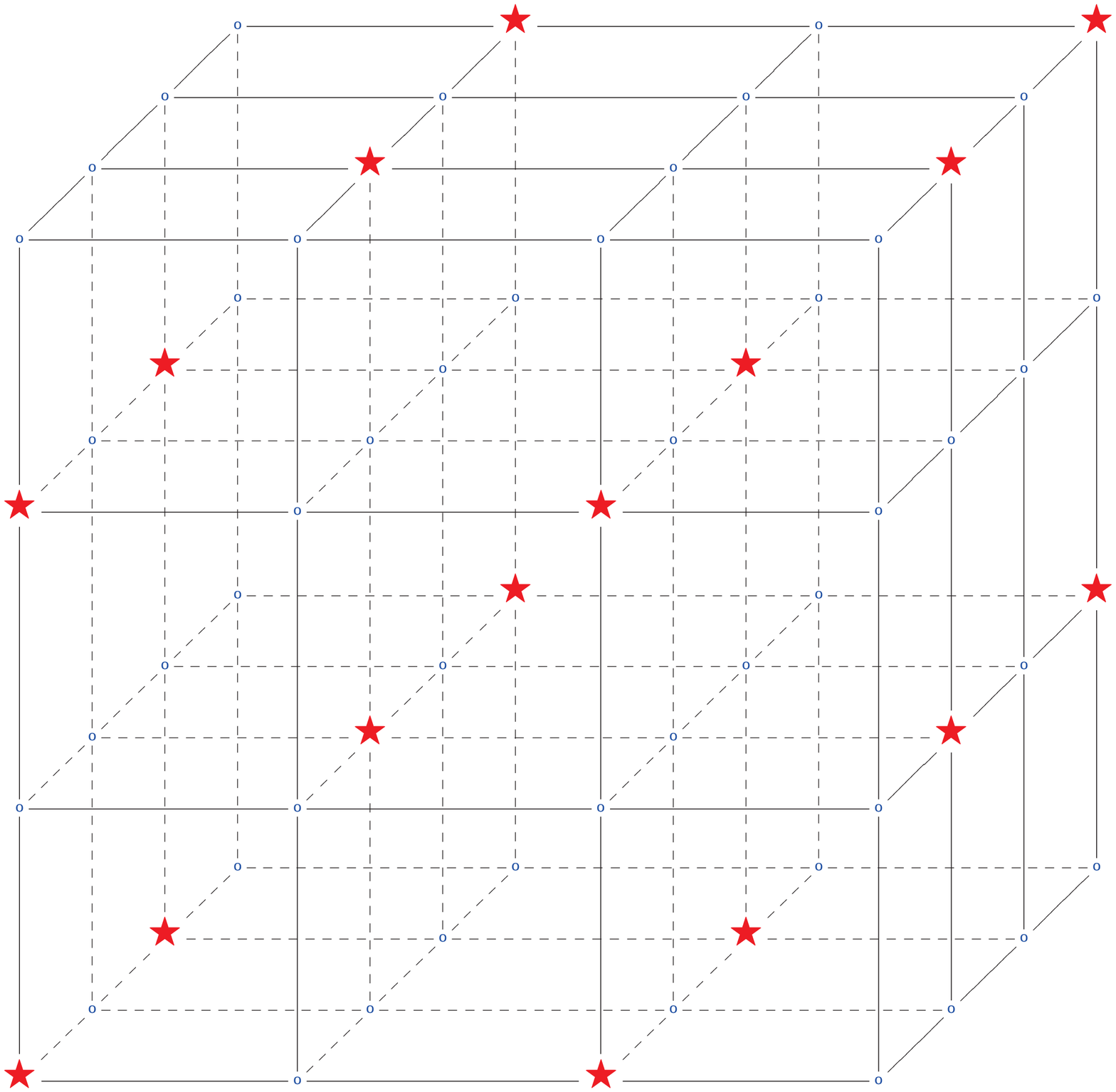}
\par\end{centering}
\caption{Example of combined mode on $4^3$ 3-D lattice. Red stars and black triangles in the left {\color{black}panel} are two subsets colored in normal symmetric coloring scheme {\color{black}$P(2,2,2,0)$}, and in the right {\color{black}panel} they are combined into one note by red stars in combined mode {\color{black}scheme $P(2,2,2,2)$}. Note that the minimal distance changes {\color{black}from $2$ to $\sqrt{3}$}. }
\label{Fig:3D_combine}
\end{figure}

We can also combine two {\color{black}subsets} into a new subset which include two sites with the space-time
offset equals $(r/2,r/2,r/2,r/2)$, then the minimal distance remains unchanged, that is
$r_{\mathrm{min}}^{\prime}=r$ for the new subset. We call this mode by Combine Mode, while
number of sites doubles and errors are increased due to more nearest neighbors are introduced.
Note that this combined mode only works on 4-D lattices , and $r$ should be even.
Fig. \ref{Fig:2D_combine} and Fig. \ref{Fig:3D_combine} show examples of
Combine Mode in 2D and 3D situations, while $r_{\mathrm{min}}=r/\sqrt{2}$ and
{\color{black}$r_{\mathrm{min}}=\sqrt{3}r/2$} respectively after combination.

Considering {\color{black}this} on a lattice $L(n_{x},n_{y},n_{z},n_{t})$.
{\color{black}Firstly} choose a symmetric coloring scheme $P\left(m_{x,}m_{y},m_{z},m_{t},\mathrm{mode}\right)$,
which means $m_{\mu}$ sites are marked with same color in $\mu$ direction
of the lattice, and the mode label can be {\color{black}$0,1,2$} which respectively means Normal,
Split and Combine Mode. {\color{black}Then pick} a site $X(x_{1},x_{1},x_{3},x_{4})$ as
a seed site with $x_{\mu}$ denotes its coordinate {\color{black}on} the lattice,
we can construct a SMP source $\phi_{P}$ in scheme $P\left(m_{x,}m_{y},m_{z},m_{t},\mathrm{mode}\right)$:
\begin{equation}
\phi_{P}\left(S\left(X,P\right),\alpha,a\right)=\sum_{y\in S\left(X,P\right)}\psi\left(y,\alpha,a\right),
\end{equation}
{where $S\left(X,P\right)$ denotes the site subset that including all sites with the same color of site
$X$ by coloring scheme $P\left(m_{x,}m_{y},m_{z},m_{t},\mathrm{mode}\right)$.}

The topological charge
density on any site $x\in S\left(X,P\right)$ can be calculated by
this combined source:
\begin{equation}
q^{\mathrm{approx}}(x)=\sum_{\alpha,a}\psi(x,\alpha,a)\tilde{D}_{\mathrm{ov}}\phi_{_{P}}\left(S\left(X,P\right),\alpha,a\right)=q(x)+\sum_{\alpha,a}\sum_{y\in S(X,P)}^{y\ne x}\tilde{D}_{\mathrm{ov}}(x,\alpha,a;\ y,\alpha,a),\label{eq:qpx}
\end{equation}
the second term sums the corresponding off-diagonal components of
$D_{\mathrm{ov}}$, which satisfy Eq. (\ref{eq:Dlocal}).

Now the whole lattice sites can be divided into $N^{\mathrm{SMP}}$
site sets, with $m_{x}m_{y}m_{z}m_{t}$ sites in each set for $\mathrm{mode}=0$
or $\frac{1}{2}m_{x}m_{y}m_{z}m_{t}$ sites for $\mathrm{mode}=1$,
$2\times m_{x}m_{y}m_{z}m_{t}$ sites for $\mathrm{mode}=2$.
Each site on the lattice should belong to and only belong to one set
in a specific scheme $P$. SMP sources are constructed from
these symmetric site sets that only distinct from each other with
some units of coordinate shift. In this case all SMP sources include
the same number of sites for a specific $P$, which leads to less sources needed
compared with Greedy Multicoloring Algorithm for the same minimal
Euclid distance $\left|x-y\right|$. Summing
Eq. (\ref{eq:qpx}) over the lattice volume, the total introduced
systematic error of $Q$ will be:
\begin{eqnarray}
R_{P} & = & Q^{\mathrm{approx}}-Q=\sum_{x\in V}\sum_{\alpha,a}\sum_{y\in S(x,P)}^{y\ne x}\tilde{D}_{\mathrm{ov}}^{-1}(x,\alpha,a;\ y,\alpha,a)\nonumber \\
 & \le & 6N_{L}C\sum_{y\in S(x,P)}^{y\ne x}\exp\left(-\gamma\left|x-y\right|\right),\label{eq:Rp}
\end{eqnarray}
where Eq. (\ref{eq:Dlocal}) is used and the sum over color and Dirac
indexes results a factor of $12$. In practice the {\color{black}errors are} far smaller
than its upper bound. Define the minimal distance $r_{\mathrm{min}}^{P}$
for a SMP scheme $P$, which is the minimal Euclid distance of
any two sites in a site set $S\left(X,P\right)$ based on arbitrary
seed site $X$:
\begin{equation}
r_{\mathrm{min}}^{P}=\mathrm{min}\left\{ \left|x-y\right|,\forall x,y\in S\left(X,P\right),\ x\ne y\right\} ,
\end{equation}
we believe that $\left|x-y\right|=r_{\mathrm{min}}^{P}$ terms will
dominate in Eq. (\ref{eq:Rp}).

For the case of a spatial symmetric lattice $L\left(n_{s},n_{s},n_{s},n_{t}\right)$,
the SMP scheme can be chosen as $P\left(\frac{n_{s}}{r},\frac{n_{s}}{r},\frac{n_{s}}{r},\frac{n_{t}}{r},\mathrm{mode}\right)$
so that
\begin{equation}
r_{\mathrm{min}}=\begin{cases}
r, & \mathrm{mode}=0,2\\
r\sqrt{2}, & \mathrm{mode}=1.
\end{cases}
\end{equation}
Therefore the choice of $r_{\mathrm{min}}$ should be one of the common
factors of $n_{s}$ and $n_{t}$, or multiplied by $\sqrt{2}$ when
$\mathrm{mode}=1$, such as $r_{\mathrm{min}}=2$
in the left panel of {\color{black}Fig. \ref{Fig:2D_split}}, and $r_{\mathrm{min}}=2\sqrt{2}$
in the right panel. We {\color{black}chose} appropriate $r_{\mathrm{min}}$
and SMP schemes for different lattices in this way.

\section{Lattice setup and results}

We calculated topological charge density and the total topological
charge for pure gauge lattice configurations with Iwasaki
gauge action~\cite{AliKhan:2000iv}:
\begin{equation}
S_{\mathrm{IA}}=\beta c_{1}\left\{ \Omega_{p}+r_{\beta}\Omega_{r}\right\}
\end{equation}
where $\beta$ is the bare coupling constant, $\Omega_{C}=\sum\frac{1}{3}\mathrm{ReTr}\left(1-W_{C}\right)$
is the real part of the trace over all Wilson loop $W_{C}$ with closed
contour $C$, while subscripts $p$ and $r$ refer to plaquette and
$2\times1$ rectangle respectively. $c_{1}=3.648$, $r_{\beta}=-0.09073465$
in Iwasaki action which is an $\mathcal{O}\left(a^{2}\right)$ improved
gauge field action.

\begin{figure}[h t b]
\begin{centering}
\includegraphics[scale=0.35]{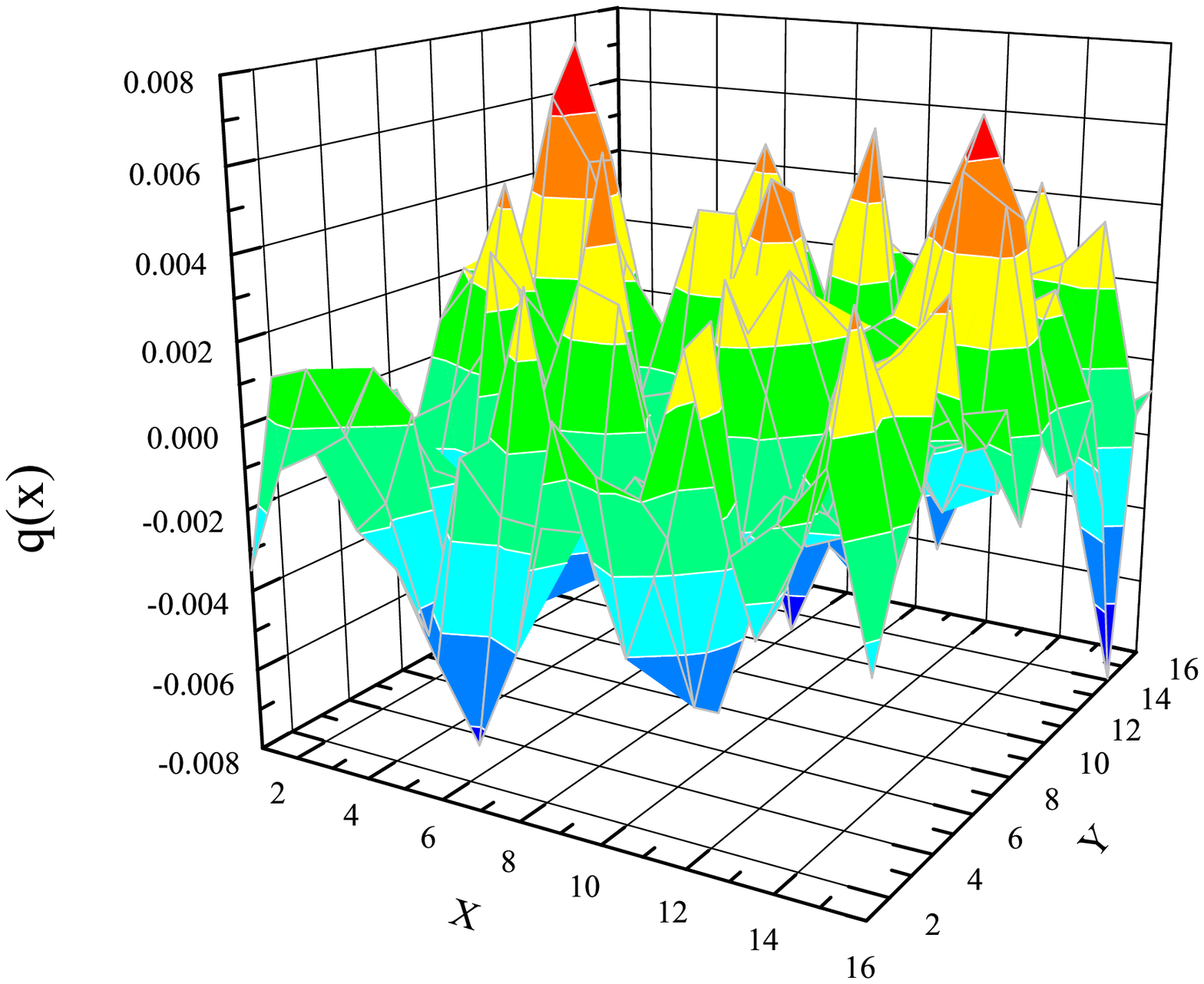}\includegraphics[scale=0.35]{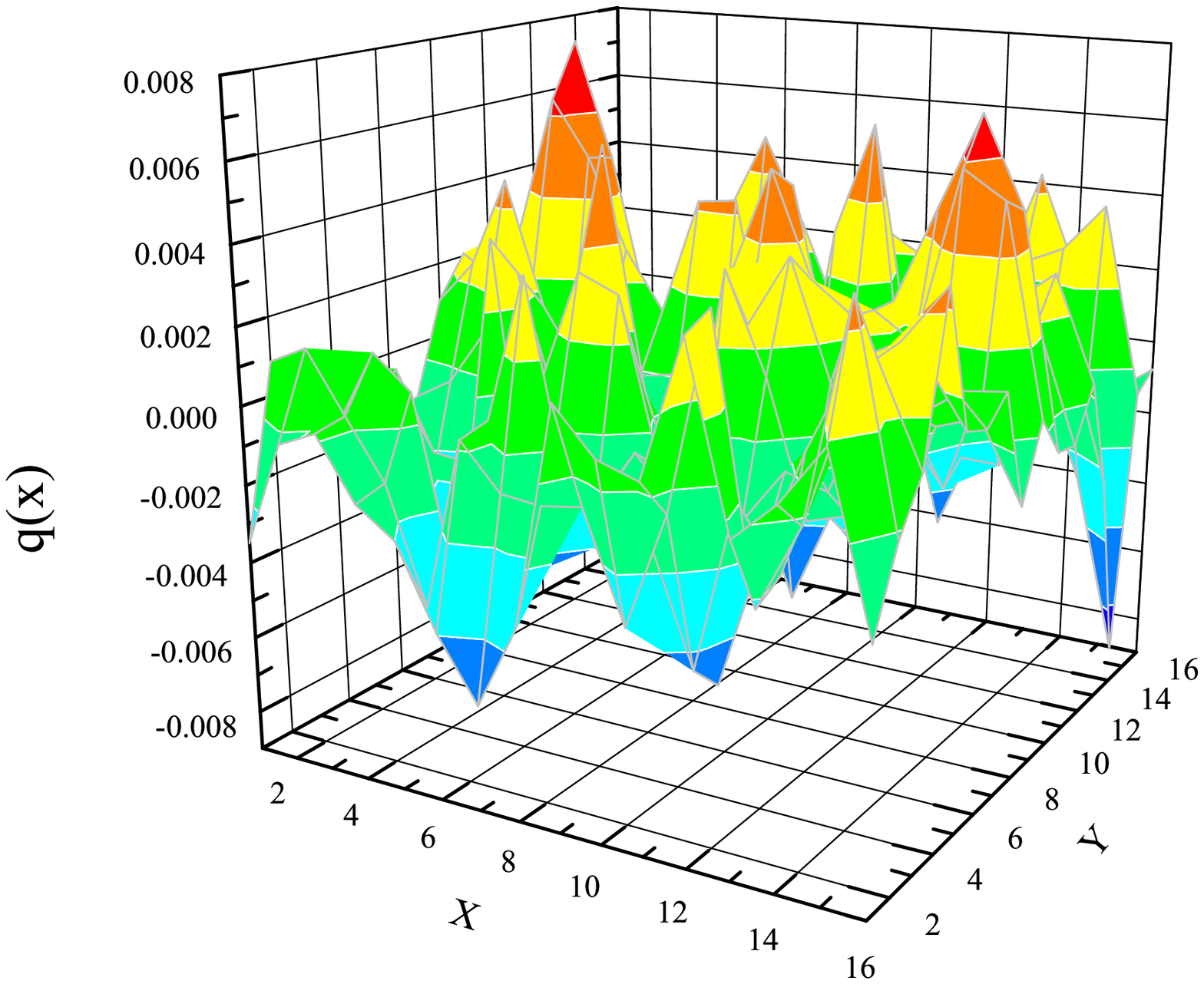}
\par\end{centering}

\caption{Topological charge density in a X-Y plane, by tracing $D_{\mathrm{ov}}$
with point sources (left) and SMP sources with $r_{\mathrm{min}}=4\sqrt{2}$
(right) on a $16^{4}$ configuration.}
\label{Fig:qx_1p_mp}
\end{figure}

Firstly we calculated topological charge density exactly with point
sources on a $16{}^{4}$ lattice and compared it to that using SMP
sources with $r_{\mathrm{min}}=4\sqrt{2}$. The results of topological
charge density in $X-Y$ plane with $t=1$ and $z=1$ are shown in
Fig. \ref{Fig:qx_1p_mp}. It's obvious that there are quite similar
peaks and valleys between the results by using point sources and SMP
sources, which indicates the topological structure of the QCD
vacuum is well reserved in the SMP method.

A series of gauge field configurations were then generated with different
lattice spacing $a$ and different lattice volume. Topological charge
density for each configuration was evaluated by SMP sources in different
schemes, see Table \ref{tab:Lattice-setup}, where the SMP scheme
$P\left(m,m,m,m,\mathrm{mode}\right)$ is written as $\left(m,\mathrm{mode}\right)$
for short, and the lattice $L\left(n,n,n,n\right)$ is written as $n^{4}$. The
number of point sources $N^{\mathrm{PT}}$ and SMP sources
$N^{\mathrm{SMP}}$ are presented in Table \ref{tab:Required-CG-solution},
which are also the number of linear Eq. (\ref{eq:Mx=00003Db}) that we had
to solve by CG algorithm. It's shown that the cost of calculating
the topological charge density was sharply reduced for smaller $r_{\mathrm{min}}$
and larger lattices.

\begin{table}
\begin{centering}
\begin{tabular}{|c|c|c|c|c|}
\hline
\multirow{2}{*}{lattice} & \multirow{2}{*}{$a/\mathrm{fm}$} & \multirow{2}{*}{No. of config.} & \multirow{2}{*}{SMP scheme} & \multirow{2}{*}{$r_{\mathrm{min}}$}\tabularnewline
 &  &  &  & \tabularnewline
\hline
\hline
\multirow{2}{*}{$12^{4}$} & 0.1 & 20 & \multirow{2}{*}{$\left(6,0\right)$$\left(6,1\right)$$\left(4,0\right)$$\left(4,1\right)$$\left(3,2\right)$$\left(3,0\right)$$\left(2,0\right)$} & \multirow{2}{*}{$2$, $2\sqrt{2}$, $3$, $3\sqrt{2}$, $4$, $4$, $6$}\tabularnewline
\cline{2-3}
 & 0.133 & 20 &  & \tabularnewline
\hline
\multirow{2}{*}{$16^{4}$} & \multirow{2}{*}{0.1} & \multirow{2}{*}{20} & \multirow{2}{*}{$\left(8,0\right)$$\left(8,1\right)$$\left(4,2\right)$$\left(4,0\right)$$\left(4,1\right)$$\left(2,0\right)$} & \multirow{2}{*}{$2$, $2\sqrt{2}$, $4$, $4$, $4\sqrt{2}$, $8$}\tabularnewline
 &  &  &  & \tabularnewline
\hline
\multirow{2}{*}{$24^{4}$} & \multirow{2}{*}{0.1} & \multirow{2}{*}{20} & \multirow{2}{*}{$\left(12,0\right)$$\left(12,1\right)$$\left(8,0\right)$$\left(8,1\right)$$\left(6,2\right)$$\left(6,0\right)$} & \multirow{2}{*}{$2$, $2\sqrt{2}$, $3$, $3\sqrt{2}$, $4$, $4$}\tabularnewline
 &  &  &  & \tabularnewline
\hline
\multirow{2}{*}{$32^{4}$} & \multirow{2}{*}{0.1} & \multirow{2}{*}{20} & \multirow{2}{*}{$\left(16,0\right)$$\left(16,1\right)$$\left(8,2\right)$$\left(8,0\right)$} & \multirow{2}{*}{$2$, $2\sqrt{2}$, $4$, $4$}\tabularnewline
 &  &  &  & \tabularnewline
\hline
\end{tabular}
\par\end{centering}

\caption{Lattice setup and multi-probing scheme.\label{tab:Lattice-setup}
Symmetric schemes $P\left(m,m,m,m,\mathrm{mode}\right)$ are written
as $\left(m,\mathrm{mode}\right)$ for short.}
\end{table}

\begin{table}
\begin{centering}
\begin{tabular}{|c|c|c|}
\hline
\multirow{2}{*}{lattice} & \multirow{2}{*}{$N^{\mathrm{SMP}}$} & \multirow{2}{*}{$N^{\mathrm{SMP}}/N^{\mathrm{PT}}$}\tabularnewline
 &  & \tabularnewline
\hline
\hline
\multirow{2}{*}{$12^{4}$} & \multirow{2}{*}{192, 384, 972, 1944, 1536, 3072, 15552} & \multirow{2}{*}{$\frac{1}{1296}$, $\frac{1}{648}$, $\frac{1}{256}$, $\frac{1}{128}$,
$\frac{1}{162}$, $\frac{1}{81}$, $\frac{1}{16}$}\tabularnewline
 &  & \tabularnewline
\hline
\multirow{2}{*}{$16^{4}$} & \multirow{2}{*}{192, 384, 1536, 3072, 6144, 49152} & \multirow{2}{*}{$\frac{1}{4096}$, $\frac{1}{2048}$, $\frac{1}{512}$, $\frac{1}{256}$, $\frac{1}{128}$,
$\frac{1}{16}$}\tabularnewline
 &  & \tabularnewline
\hline
\multirow{2}{*}{$24^{4}$} & \multirow{2}{*}{192, 384, 972, 1944, 1536, 3072} & \multirow{2}{*}{$\frac{1}{20736}$, $\frac{1}{10368}$, $\frac{1}{4096}$, $\frac{1}{2048}$, $\frac{1}{2592}$,
$\frac{1}{1296}$}\tabularnewline
 &  & \tabularnewline
\hline
\multirow{2}{*}{$32^{4}$} & \multirow{2}{*}{192, 384, 1536, 3072} & \multirow{2}{*}{$\frac{1}{65536}$, $\frac{1}{32768}$, $\frac{1}{8192}$, $\frac{1}{4096}$}\tabularnewline
 &  & \tabularnewline
\hline
\end{tabular}
\par\end{centering}

\caption{Number of sources $N^{\mathrm{SMP}}$ required in different SMP scheme,
and the ratio to $N^{\mathrm{PT}}$ referring to {\color{black}the number of} point sources.\label{tab:Required-CG-solution}}
\end{table}

\begin{figure}[h t b]
\begin{centering}
\includegraphics[scale=0.4]{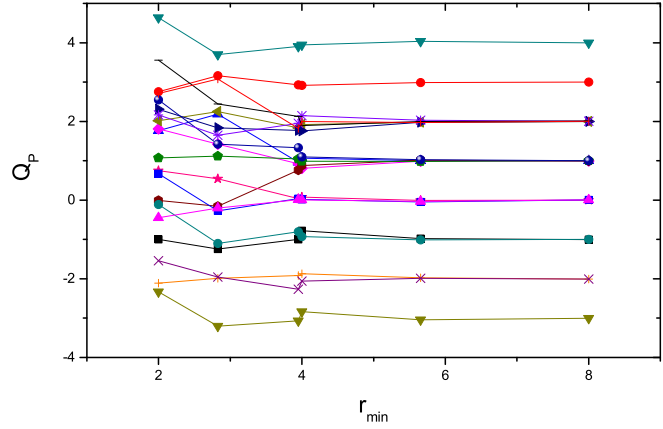}
\par\end{centering}

\caption{The series of $Q_{P}$ with different $r_{\mathrm{min}}$ for each
configuration {\color{black}on $16^{4}$} lattice with $a=0.1\mathrm{fm}$.
When $r_{\mathrm{min}}\geq4$, $Q_{P}$
is very close to an integer value. Exact integer number of the topological
charge {\color{black}were} extracted since the systematic errors should be sharply suppressed
as $r_{\mathrm{min}}$ increases.}
\label{Fig:16a.1Q}
\end{figure}

\begin{figure}[h t b]
\begin{centering}
\includegraphics[scale=0.3]{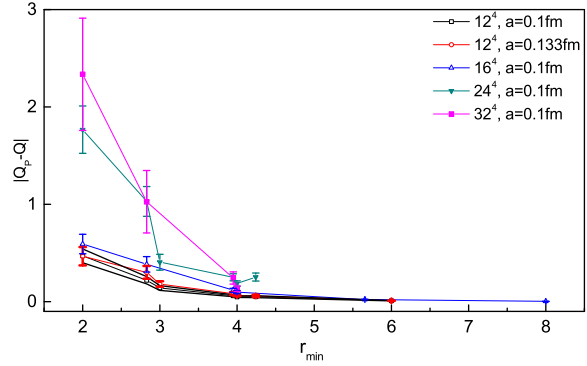}
\par\end{centering}

\caption{The average absolute error of $Q$ vs. $r_{\mathrm{min}}$ for all
lattice setting. For $r_{\min}\ge4$, the errors are small.}
\label{Fig:all-r}
\end{figure}

Since $Q_{P}$ (topological charge calculated in
scheme $P$) should approach an integer and the deviation
of $Q_{P}$ to its exact value will be always smaller than 0.1 as
long as $r_{\mathrm{min}}$ is large enough, we {\color{black}extracted} the topological
charge $Q$ and absolute value of systematic error $\left|Q_{P}-Q\right|$ from
the series of results with different $r_{\mathrm{min}}$ for each
configuration, like an example in Fig. \ref{Fig:16a.1Q}. The average
errors are shown in Fig. \ref{Fig:all-r}, which decrease sharply
as $r_{\mathrm{min}}$ increases. In general, the average absolute
value of systematic error of $Q_{P}$ and its variance {\color{black}became} small enough when $r_{\mathrm{min}}\ge4$.
One may notices that the error with $r_{\mathrm{min}}=3\sqrt{2}$
on $24^{4}$ lattice {\color{black}was} larger than that of $r_{\mathrm{min}}=4$,
which may {\color{black}come} from more nearest neighbor terms (satisfying
$\left|x-y\right|=r_{\mathrm{min}}$ in Eq. (\ref{eq:Rp}))
in $\mathrm{mode}=1$ case than $\mathrm{mode}=0$. {\color{black}There are
24 nearest neighbors for each site in the former mode while only 8 nearest neighbors in the latter mode}.

\begin{figure}[h t b]
\begin{centering}
\includegraphics[scale=0.3]{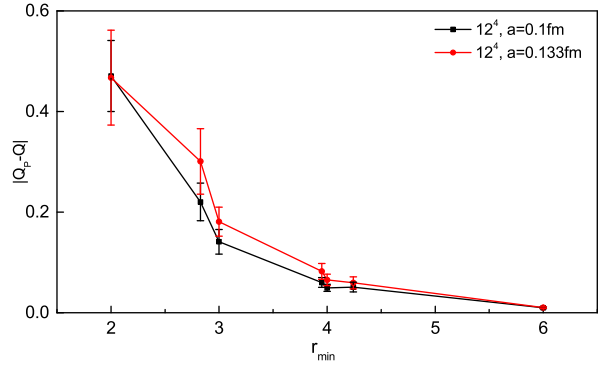}
\par\end{centering}

\caption{Comparing the average absolute error of $Q_{P}$ vs. $r_{\mathrm{min}}$
on $12^{4}$ lattice {\color{black}with} different $a$. For the same $r_{\mathrm{min}}$,
the finer lattice with smaller $a$ leads to smaller deviation of
$Q$ to integer.}
\label{Fig:x12cpr}
\end{figure}

\begin{figure}[h t b]
\begin{centering}
\includegraphics[scale=0.3]{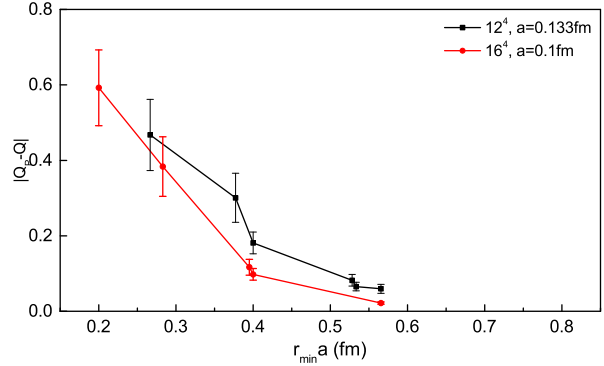}
\par\end{centering}

\caption{Comparing the average absolute error of $Q_{P}$ vs. physical distance
$r_{\mathrm{min}}a$ on $V=\left(1.6\mathrm{fm}\right)^{4}$ lattice
between different $a$. In physical unit, the errors are also smaller
on finer lattice with the same volume.}
\label{Fig:L16cpr}
\end{figure}

\begin{figure}[h t b]
\begin{centering}
\includegraphics[scale=0.3]{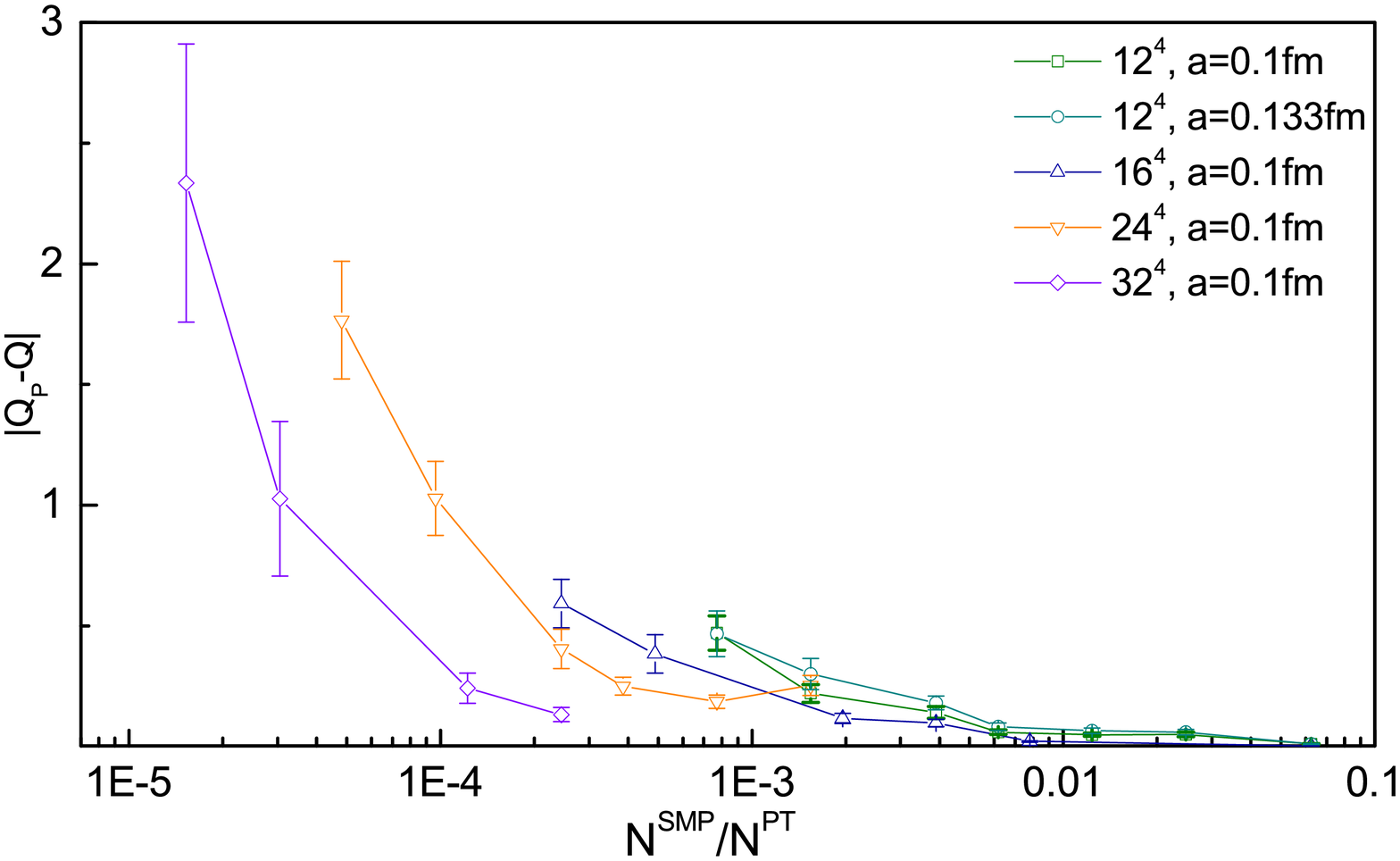}
\par\end{centering}

\caption{The errors versus the efficiency for all lattices. Larger lattice
has higher efficiency.}
\label{Fig:all_npn}
\end{figure}

{\color{black}Considering} the same lattice with different lattice spacing $a$, Fig.
\ref{Fig:x12cpr} shows that finer lattice spacing $a$ leads to smaller
errors. It could also be concluded from Fig. \ref{Fig:L16cpr}, which
compares two lattices with the same physical volume. We finally present
the errors versus efficiency in Fig. \ref{Fig:all_npn}, which shows
that for larger lattices, the calculation of trace by SMP method {\color{black}is
more efficient} to achieve sufficient accuracy.

\begin{table}
\begin{centering}
\begin{tabular}{|c|c|c|c|c||c|c|c|c|c|}
\hline
$L$ & $a/\mathrm{fm}$ & $P$ & $N^{\mathrm{SMP}}$ & $\left\langle \left|Q_{P}-Q\right|\right\rangle $ & $L$ & $a/\mathrm{fm}$ & $P$ & $N^{\mathrm{SMP}}$ & $\left\langle \left|Q_{P}-Q\right|\right\rangle $\tabularnewline
\hline
\hline
\multirow{14}{*}{$12^{4}$} & \multirow{7}{*}{$0.1$} & $\left(6,0\right)$ & 192 & $0.471\left(71\right)$ & \multirow{5}{*}{$16^{4}$} & \multirow{5}{*}{$0.1$} & $\left(8,1\right)$ & 384 & $0.384\left(79\right)$\tabularnewline
\cline{3-5} \cline{8-10}
 &  & $\left(6,1\right)$ & 384 & $0.220\left(37\right)$ &  &  & $\left(4,2\right)$ & 1536 & $0.0117\left(21\right)$\tabularnewline
\cline{3-5} \cline{8-10}
 &  & $\left(4,0\right)$ & 972 & $0.141\left(24\right)$ &  &  & $\left(4,0\right)$ & 3072 & $0.098\left(15\right)$\tabularnewline
\cline{3-5} \cline{8-10}
 &  & $\left(4,1\right)$ & 1944 & $0.051\left(10\right)$ &  &  & $\left(4,1\right)$ & 6144 & $0.022\left(3\right)$\tabularnewline
\cline{3-5} \cline{8-10}
 &  & $\left(3,2\right)$ & 1536 & $0.060\left(10\right)$ &  &  & $\left(2,0\right)$ & 49152 & $0.003\left(1\right)$\tabularnewline
\cline{3-5} \cline{6-10}
 &  & $\left(3,0\right)$ & 3072 & $0.049\left(7\right)$ & \multirow{6}{*}{$24^{4}$} & \multirow{6}{*}{$0.1$} & $\left(12,0\right)$ & 192 & $1.767\left(244\right)$\tabularnewline
\cline{3-5} \cline{8-10}
 &  & $\left(2,0\right)$ & 15552 & $0.010\left(2\right)$ &  &  & $\left(12,1\right)$ & 384 & $1.029\left(153\right)$\tabularnewline
\cline{2-5} \cline{8-10}
 & \multirow{7}{*}{$0.133$} & $\left(6,0\right)$ & 192 & $0.467\left(94\right)$ &  &  & $\left(8,0\right)$ & 972 & $0.406\left(81\right)$\tabularnewline
\cline{3-5} \cline{8-10}
 &  & $\left(6,1\right)$ & 384 & $0.301\left(65\right)$ &  &  & $\left(8,1\right)$ & 1944 & $0.253\left(41\right)$\tabularnewline
\cline{3-5} \cline{8-10}
 &  & $\left(4,0\right)$ & 972 & $0.181\left(29\right)$ &  &  & $\left(6,2\right)$ & 1536 & $0.250\left(36\right)$\tabularnewline
\cline{3-5} \cline{8-10}
 &  & $\left(4,1\right)$ & 1944 & $0.059\left(12\right)$ &  &  & $\left(6,0\right)$ & 3072 & $0.186\left(27\right)$\tabularnewline
\cline{3-5} \cline{6-10}
 &  & $\left(3,2\right)$ & 1536 & $0.082\left(16\right)$ & \multirow{4}{*}{$32^{4}$} & \multirow{4}{*}{$0.1$} & $\left(16,0\right)$ & 192 & $2.335\left(576\right)$\tabularnewline
\cline{3-5} \cline{8-10}
 &  & $\left(3,0\right)$ & 3072 & $0.065\left(11\right)$ &  &  & $\left(16,1\right)$ & 384 & $1.026\left(320\right)$\tabularnewline
\cline{3-5} \cline{8-10}
 &  & $\left(2,0\right)$ & 15552 & $0.010\left(2\right)$ &  &  & $\left(8,2\right)$ & 1536 & $0.242\left(63\right)$\tabularnewline
\cline{1-5} \cline{8-10}
$16^{4}$ & $0.1$ & $\left(8,0\right)$ & 192 & $0.592\left(101\right)$ &  &  & $\left(8,0\right)$ & 3072 & $0.133\left(29\right)$\tabularnewline
\hline
\end{tabular}
\par\end{centering}

\caption{{The average of absolute systematic error $\left\langle \left|Q_{P}-Q\right|\right\rangle $
for different SMP scheme $P$.}\label{tab:systematic error}}
\end{table}

{
All results of $\left\langle \left|Q_{P}-Q\right|\right\rangle$ are presented in Table \ref{tab:systematic error}.
We can see for example, when $N^{\mathrm{SMP}}=3072$ the system error increases slowly as the lattice becomes larger.
This could be understood since the number of non-zero off-diagonal elements in SMP sources is proportional to $N_L$, while
$N^{\mathrm{SMP}}=3072$ is fixed.
We can expect to apply this method on larger lattice with fixed $N^{\mathrm{SMP}}$, while $N_{\mathrm{PT}}$ is
proportional to {\color{black}the lattice volumes} $N_L$.}

{\color{black}
There is another method to calculate topological charge by calculating
a few low modes of chiral Dirac operator, counting the number of zero
modes and applying the index theorem. However, to estimate the topological
charge density $q\left(x\right)$ one must calculate more low modes
by applying low mode expansion \cite{horvath2003local}. We roughly
compared the cost of SMP method and the calculation of low modes of
the overlap operator on $12^{4}$ lattices. We considered $N^{\mathrm{SMP}}=3072$
with scheme $P\left(3,3,3,3,0\right)$, $r_{\mathrm{min}}=4$ in SMP
method, and calculated the low modes of overlap operator by Arnordi
algorithm with the number of low modes $N^{\mathrm{mode}}=100,\ 200$.
The results showed that when $N^{\mathrm{mode}}=100$ , low modes
calculation costs about $18\%$ less than SMP method, while $N^{\mathrm{mode}}=200$
it costs $28\%$ more than SMP method. However, since the number of
zero modes is always an integer, the systematic errors of $q\left(x\right)$
from low mode expansion could hardly be estimated, while in SMP method
the systematic errors of $Q$ and $q\left(x\right)$ are related.
The comparison of $q\left(x\right)$ from SMP method and low mode
expansion will be a interesting topic in our further studies.
}

\section{Conclusion}

Topological charge $Q$ and density $q\left(x\right)$ were calculated
from the trace of overlap Dirac operator employing symmetric multi-probing
sources on pure gauge configurations. The systematic error were suppressed
when $r_{\mathrm{min}}$ increased as expected, due to the locality of
overlap Dirac operator. Thus the estimated topological charge were
close enough to an integer while $r_{\mathrm{min}}\ge4$ or $r_{\mathrm{min}}\ge4\sqrt{2}$,
in which case the number of required sources $N^{\mathrm{SMP}}=12r_{\mathrm{min}}^{4}\ \mathrm{or}\ 24r_{\mathrm{min}}^{4}$
comparing to that with point sources $N^{\mathrm{PT}}=12N_{L}$. Notice
that $N^{\mathrm{SMP}}$ only depends on the chosen $r_{\mathrm{min}}$
and is independent of $N_{L}$. Furthermore, we
found that the introduced systematic {\color{black}errors} of SMP method for the same $r_{\mathrm{min}}$
grew slowly with larger $N_{L}$, which {\color{black}meant} the time to estimate
the trace is much less dependent on the lattice volumes $N_{L}$. Thus
it will be much more economical to employ SMP
method on larger lattices. Since the introduced error of the trace
comes from the off-diagonal elements of the matrix, we can expect
better performance of this method by employing some techniques to
suppress the off-diagonal elements of $D_{\mathrm{ov}}$ in future
work. This method is also supposed to work well when {\color{black}dealing} with ultra-local
matrix such as Wilson Dirac operator $D_{\mathrm{W}}$, {\color{black}and} $\mathrm{Tr}D_{\mathrm{W}}^{-1}$
is very important and quite expensive to {\color{black}determinate} exactly in lattice
QCD.

{
As mentioned above, we developed symmetric scheme with 3 modes in coloring algorithm and evaluated the topological charge
with SMP method. Results of average absolute systematic error are presented, as well as the efficiency of SMP method
compared to point source method. Potential applications of SMP method are expected in calculating the trace of the inverse of
any large sparse matrix with locality.
}

\section*{Acknowledgments}
This work mainly ran on Tianhe-1A at NSCC in Tianjin. This work
is supported by the National Natural Science Foundation of
China (NSFC) under the project No.11335001, No.11275169.

\section{References}

\bibliographystyle{unsrt}
\bibliography{refer}


\clearpage
\end{CJK*}
\end{document}